\documentclass[twocolumn,usenatbib]{mnras}
\pdfoutput=1 
\usepackage{amsmath,amstext}
\usepackage[T1]{fontenc}
\usepackage{amssymb}
\usepackage{amsmath}
\input{epsf}
\usepackage{graphicx}

\usepackage{color}

\newcommand{\n}{{\mathrm{n}}}
\newcommand{\p}{{\mathrm{c}}}

\newcommand{\x}{\mathrm{x}}
\newcommand{\y}{\mathrm{y}}
\newcommand{\vv}{\mathrm{v}}
\newcommand{\R}{\mathcal{R}}

\def\be{\begin{equation}}
\def\ee{\end{equation}}
\def\beq{\begin{eqnarray}}
\def\eeq{\end{eqnarray}}

\title[Turbulent, pinned superfluids and glitch recoveries]{Turbulent, pinned superfluids in neutron stars and pulsar glitch recoveries.}
\author[B.~Haskell, D.~Antonopoulou \& C.~Barenghi ]{B.~Haskell$^{1}$, D.~Antonopoulou$^{1}$ \& C.~Barenghi$^{2}$ \\
$^{1}$ Nicolaus Copernicus Astronomical Center, Polish Academy of Sciences, ul. Bartycka 18, 00-716 Warsaw, Poland\\
$^{2}$ School of Mathematics, Statistics and Physics, Newcastle University, Newcastle upon Tyne NE1 7RU, 
United Kingdom}

\begin{document}
\maketitle

\begin{abstract}

Pulsar glitches offer an insight into the dynamics of superfluids in the high density interior of a neutron star. To model these phenomena, however, one needs to have an understanding of the dynamics of a turbulent array of superfluid vortices moving through a pinning lattice. In this paper we develop a theoretical approach to describe vortex mediated mutual friction in a pinned, turbulent and rotating superfluid. Our model is then applied to the study of the post glitch rotational evolution in the Vela pulsar and in PSR J0537-6910. We show that in both cases a turbulent model fits the evolution of the spin frequency derivative better than a laminar one. We also predict that the second derivative of the frequency after a glitch should be correlated with the waiting time since the previous glitch, which we find to be consistent with observational data for these pulsars. The main conclusion of this paper is that in the post-glitch rotational evolution of these two pulsars we are most likely observing the response to the glitch of a pinned turbulent region of the star (possibly the crust) and not the laminar response of a regular straight vortex array.

 \end{abstract}

\section{Introduction} 

The dynamics of quantized vorticity in superfluid helium II and cold atomic gases has been studied extensively in the past decades, both theoretically and experimentally \citep{DonnellyBook, DB2001, Rev14}. In particular, studies have focused mostly on two regimes: rotating superfluids, in which an array of straight parallel vortices is generated, and counterflow, in which the presence of a heat flux leads to a turbulent, isotropic, vortex tangle. There is, however, another system of physical interest in which both rotation and counterflow are likely to be important, and the dynamics of a polarized, turbulent, quantized vortex tangle can be studied. This is the interior of a neutron star.

Neutron stars are an extraordinary laboratory in which to study fundamental physics in extreme conditions that cannot be reproduced by terrestrial experiments. Not only are these stars incredibly dense -- as with a mass comparable to that of the sun compressed into a roughly $10$ km radius, their central densities can be several times nuclear saturation density -- but they are also `cold'. The internal temperatures of mature neutron stars are, in fact, well below the Fermi temperature at such densities: for a typical star the thermal energy will be of order $1$ keV, while the Fermi energy is of order $50$ MeV. The neutrons in the crust pair, forming a superfluid, while in the core both neutrons and protons can pair, and are respectively superfluid and superconducting (see \citet{HasSed} for a recent review).

Superfluidity substantially alters the dynamics of the system, by introducing additional degrees of freedom and the possibility of a relative flow between the superfluid components and the `normal' part of the star, such as the solid crust {whose rotation we track in pulsars by observing the electromagnetic emission}. One of the most striking observational effects that is thought to be associated with superfluidity is the glitching behaviour {of young pulsars}. Glitches are sudden increases in frequency, {typically} instantaneous to the accuracy of the data (although see \citet{Ashton, Montoli} for a discussion of the rise-time of glitches in the Vela pulsar). To date they have been {discovered} in {nearly two hundred neutron stars} \citep{els+11}. Their origin is still debated \citep{hm15}, but most models assume the presence of a large scale neutron superfluid component whose vortices are pinned to nuclear clusters in the crust. If this is the case, the superfluid cannot spin-down and develops a rotational lag with respect to the normal part of the crust, storing angular momentum. Once the lag becomes too large, hydrodynamical lift forces will unpin the vortices, leading to a rapid exchange in angular momentum and an observable glitch \citep{ai75}.

Observations of glitches, and in particular of the post-glitch {rotational relaxation}, offer an insight into the dynamics of superfluids at high densities \citep{Alpar84a, as88, acp89, nonlinear}. The main source of coupling between a normal fluid an a superfluid is the so-called mutual friction, which is mediated by the vortex array and thus depends significantly on its configuration, e.g. whether it is rectilinear or a turbulent tangle \citep{GM, HV56, BK}. 
To date very few experiments have dealt with turbulence polarized by rotation \citep{YG78, Swanson83, Finne03}, and only a limited number of studies {investigate} mutual friction for such systems \citep{ SideryTURB, Sciacca08, Jou11, Mongiovi}. Neutron star interiors are, however, thought to be exactly in such a polarized turbulent regime \citep{SideryTURB}, where differences in velocity between the normal and superfluid components can lead to instabilities \citep{Sidery08, Glamp09, Hogg13, Khomenkoinst}, and are possibly linked to the glitch trigger \citep{Peralta1, Peralta2, Mongiovi}. Furthermore, the neutron star case involves pinning in the bulk of the superfluid, with pinning sites acting as a grid that is continuously passed through the superfluid as the rotational lag develops, and it is well established from the study of terrestrial superfluids that the presence of such irregularities and pinning sites can aid the formation of vortex rings and lower the threshold for the onset of turbulence \citep{Stagg17}.

To tackle this problem we will consider the evolution of vortex lines in a neutron star setting, and analyse the case in which pinning is present in the bulk of the superfluid. We then discuss the form of the mutual friction for pinned superfluids and its possible signature in the post-glitch relaxation of the Vela pulsar and of PSR J0537-6910. Our model is compared to timing data for both these stars and the main conclusion of this paper is that a turbulent model better describes the post glitch relaxation in these pulsars. The superfluid vortices in the neutron star crust thus most likely form a turbulent tangle, and not a regular array of straight vortex lines.

\section{Multifluid equations of motion}

To model superfluids in a neutron star setting we will use the formalisms of \citet{AndComer}. 
This formalism is similar in spirit to the HVBK formalism used to describe superfluid helium \citep{HV56, BK}, as it provides the corse-grained equations of motion for two dynamical degrees of freedom, but differs from the helium case as the two degrees of freedom are not a 'superfluid' coupled to its excitations (the `normal' fluid), but rather a superfluid neutron condensate at zero temperature (labelled as $\n$) and a charge-neutral massive fluid consisting of protons and electrons locked together by electromagnetic interactions on time scales shorter than those of interest for our problem (labeled as $\p$). The evolution equations for the momentum can be written as:
\be
\left(\frac{\partial }{\partial t}+v_{\x}^{j}\nabla _j\right)\tilde{p}_{i}^{\x}+\varepsilon_\x w_{j}^{\y\x}\nabla_iv_{\x}^{j}+\nabla_i(\tilde{\mu}_\x+\Phi)=\frac{F_{i}^{\x}}{\rho_\x}\, , \label{Euler}
\ee
where $\x$ and $\y$ label the constituents, assuming $\x\neq\y$, while standard latin indices $i,j,k$ label the spatial coordinates. Summation is implied only over the spatial indices and not the constituent indices. In this notation $v_{i}^\x$ is the velocity of constituent $\x$, while $w_{i}^{\y\x}=v_i^\y-v_i^\x$ is the difference in velocities of components. $\Phi$ is the gravitational potential which enters the Poisson equation:
\be
\nabla^2\Phi=4\pi G\sum_\x \rho_\x \, ,
\ee
 $\tilde{\mu}_\x = \mu_x/m_\x$ is the chemical potential per unit mass, and $\rho_\x$ is the density of the $\x$ constituent and we make the approximation $m_\p=m_\n=m$. The momentum per unit mass $\tilde{p}_{i}^{\x}$ is:
\beq
\tilde{p}_{i}^\mathrm{x}=v_{i}^\mathrm{x}+\varepsilon_\x w_{i}^\mathrm{yx} \, , \label{Momentum}
\eeq 
where $\varepsilon_\x$ is the entrainment coefficient. The continuity equations, if we assume that no reactions take place over the timescales of interest, take the form:
\beq
\frac{\partial \rho_\x}{\partial t}+\nabla_j(\rho_\x v_{\x}^{j})=0 \, .\label{Continuo}
\eeq
The force $F_{i}^{\x}$ on the right hand side {of Eq. \ref{Euler}} will be the focus of our discussion, and is the vortex mediated mutual friction \citep{HV56}. On the hydrodynamical, corse-grained scale described by the equations of motion above, it represents an average, over the small volume of a fluid element, of individual interactions between vortices and the normal fluid on the sub-hydrodynamical scale. We recall that on the microscopic scale the superfluid is irrotational and the circulation is carried by the quantized vortices. A large scale velocity for the superfluid can thus only be obtained by averaging over many vortices, the distance between {which sets the scale above which one can treat the problem hydrodynamically, typically of the order of $l\approx 10^{-2} \nu$ cm, with $\nu=\Omega/2\pi$ the rotation frequency of the star in Hz}. Consequently, the mutual friction form depends strongly on the properties of the vortex array.
To obtain the force per unit volume that is needed in (\ref{Euler}) we start by analysing an individual vortex. If we neglect the inertia of the line and assume, for now, that the vortex is not pinned, then we must balance the drag and Magnus forces that act on it:
\be
\epsilon^{ijk}{\kappa}_j(v_k^\vv-v_k^\n)+\kappa\mathcal{R}(v_\p^i-v_\vv^i)=0\, ,\label{forza}
\ee
where $\mathcal{R}$ is a dimensionless drag parameter and $v_i^\vv$ is the velocity of the vortex line, which from (\ref{forza}) can be written explicitly as
\beq
v^i_{\vv}&=&\frac{v^i_\n}{1+\R^2}+\frac{\R^2}{1+\R^2}v^i_\p+\frac{\R}{1+\R^2}\epsilon^{ijk}\hat{\kappa}_j w^{\p\n}_k\nonumber\\
&&+\frac{\hat{\kappa}^i}{1+\R^2}\hat{\kappa}_j w_{\p\n}^j \, .
\label{cross}
\eeq
From this we obtain the force per unit length mediated by a vortex moving in the condensate as \citep{SideryMF}:
\be
 f_{i}^{\x}={\rho_\n} \mathcal{B}^{'}\epsilon_{ijk}\kappa^j w_{\x\y}^{k}+{\rho_\n} \mathcal{B}\epsilon _{ijk}\hat{\kappa}^j\epsilon^{klm}\kappa_l  w_{m}^{\x\y} \, , 
 \label{MF}
\ee
where $\kappa^i$ is the vector tangent to the vortex line of modulus $\kappa=h/2m_\n$, the quantum of circulation (a hat represents a unit vector), and
\beq
\mathcal{B}&=&\frac{\mathcal{R}}{1+\mathcal{R}^2}\, ,\\
\mathcal{B}^{'}&=&\frac{\mathcal{R}^2}{1+\mathcal{R}^2}\, .
\eeq
Note that the coefficients $\mathcal{B}$ and $\mathcal{B}^{'}$ are the equivalent of the standard parameters (generally denoted as $\alpha$ and $\alpha^{'}$) used in HVBK hydrodynamics, however there is a significant difference, as in our formulation the difference in velocity between the two massive species $w^i_{\p\n}$ appears in the force, which is not exactly the same quantity as the difference in velocity between the `superfluid' and `normal' fluid, which appears in the standard formulation for Helium \citep{Prix04}.  To obtain the force per unit volume we need to average the expression above over all the vortices in the volume of the fluid element, $\Lambda$, an average which we denote, for any quantity $\Pi$, as $\left< \Pi \right>$, and is defined as
\be
\left< \Pi \right>=\frac{1}{L\Lambda} \int \Pi d\xi \, ,
\ee
where $L$ is the vortex line density per unit volume, and $\xi$ is an arc length along the vortex line. Therefore, the force of Eq. \ref{MF} per unit volume acting on a fluid element is
\be
F_i^\x=L \left< f_i^x \right>\label{totalMF} \, .
\ee
Note that in the case of perfect pinning of the vortex to proton clusters in the crust, one has simply $v_i^\vv=v_i^c$ and the force balance equation is
\be
\epsilon^{ijk}{\kappa}_j(v_k^\vv-v_k^\n)+F_{pin}=0 \, ,
\ee
where $F_{pin}$ is the pinning force contribution. In this case there is locally no mutual friction, as there is no motion of the vortex segment with respect to the normal fluid. {For straight pinned vortices this still holds after averaging over a fluid element, but it may not be the case if vortices bend and rings can be formed}.

\section{Evolution of the vortex array}

Let us continue our analysis by considering the evolution of the vortex line length per unit volume, which we denote as $L$. In a rotating superfluid one has
\be
L=L_s=\frac{2\Omega}{\kappa} \label{rot1} \, ,
\ee
where $\Omega$ is the rotation rate (of the star in our case), and $\kappa$ the quantum of circulation. On the other hand, in thermal counterflow turbulence, and in the absence of rotation, it is well established ( see e.g. \citealt{DB2001}) that 
\be
L=L_r=\left(\frac{\mathcal{B} V}{\kappa}\right)^2 \label{rot2} \, ,
\ee
with $\mathcal{B}$ the mutual friction coefficient, which in the Helium problem depends on temperature, and $V$ the modulus of the counterflow velocity. If both rotation and counterflow are present the situation appears to be more complex, and two critical velocities appear: a lower one, identified with the threshold for the Donnelly-Glaberson instability \citep{GD1, GD2}, below which rotation dominates and the vortex length is well approximated by (\ref{rot1}), and a higher one above which one has a transition to a turbulent tangle where (\ref{rot2}) is a good approximation for the vortex length \citep{Tsubota04}. Between these two regimes the vortex lengths due to rotation and counterflow do not appear to simply add, and for high enough rotation rates and counterflow velocities, rotation adds less vortex length than expected \citep{Swanson83}. Nevertheless, for small values of the counterflow velocity $V$ this effect is less pronounced, and the two velocities approximately add. 
The nature of the superfluid flow in neutron star interiors cannot be probed directly with experiments. 
However, from theoretical grounds, turbulence is generally expected to develop as inertial driving forces are generally stronger than dissipation in the interior of the star \citep{SideryTURB}. Furthermore, in the crust of the star, pinning allows for a relative flow of the neutron superfluid with respect to the nuclear lattice, possibly with velocities of the order of $v\approx 10^5$ cm/s \citep{Sevesopin}. While this is still significantly below the speed of counter moving sound-waves (analogous to second-sound) in the neutron superfluid, which is of the order of $c_{ss}\approx 10^7$ cm/s \citep{Khomenkoinst}, it is well established from the study of terrestrial superfluids that the presence of irregular pinning sites will facilitate the formation of vortex rings and lower significantly the threshold for turbulence to develop \citep{Stagg17}. We will thus assume that we are dealing with a polarized turbulent tangle of vortices.
Furthermore, given that in the neutron star problem the difference in rotation rate between the superfluid and the normal component, $\Delta\Omega=(\Omega_\mathrm{n}-\Omega_\mathrm{c})$, is small compared to the rotation rate of the star, i.e. $\Delta\Omega\ll \Omega$, in the following {we will assume that we are in the limit where the total vortex length $L_T$ can be approximated as the sum of (\ref{rot1}) and (\ref{rot2}):
\be
L_T\sim\frac{2\Omega}{\kappa}+\left(\frac{\mathcal{B} R\Delta\Omega}{\kappa}\right)^2\label{tot} \, ,
\ee
with $R$ the stellar radius}. 
This solution can also be obtained by considering an evolution equation for $L$, of the form 
\be
\frac{dL}{dt}=\left(\frac{dL}{dt}\right)_{\mbox{formation}}-\left(\frac{dL}{dt}\right)_{\mbox{destruction}} \, ,
\label{len1}
\ee
following the approach that was first suggested by \citet{Vinen57a, Vinen57b, Vinen57c}. Different forms are possible for the formation and destruction terms in the Vinen equations, and experiments cannot currently rule out the different alternatives, especially for the case of polarized turbulence. Two forms were proposed in the presence of a rotation rate $\Omega$ and counterflow velocity $V$, namely \citet{Jou04} proposed to modify the classical Vinen equation as 
\beq
\frac{dL}{dt}&=&-\beta\kappa L^2 +\alpha_1\left[L^{1/2}-m_1 \frac{\sqrt{\Omega}}{\sqrt{\kappa}}\right] V L+ \nonumber\\
&&\beta_2\left[L^{1/2}-m_2 \frac{\sqrt{\Omega}}{\sqrt{\kappa}}\right] \sqrt{\kappa\Omega} L \, ,
\label{vinen1}
\eeq
while \citet{Sciacca08} proposed the alternative form, also acceptable on dimensional and microphysical grounds:
\be
\frac{dL}{dt}=-\beta\kappa L^2 + A_1\left[ L-\nu_1\frac{\Omega}{\kappa}\right]\frac{V^2}{\kappa}+B_1\left[ L-\nu_2\frac{\Omega}{\kappa}\right]\Omega \, ,
\label{vinen2}
\ee
where $\beta$, $\alpha_1$, $\beta_1$, $m_1$, $m_2$, $A_1$, $B_1$, $\nu_1$ and $\nu_2$ are phenomenological  parameters to be determined from experiment. Both equations can reproduce experimental results for pure counterflow turbulence, and for simple rotation. {Furthermore, in both cases, we can consider a small counterflow velocity expansion away from an equilibrium solution for pure rotation with a rate $\Omega$. If we assume that this solution minimises the vortex length per unit volume, and thus take the first term in the expansion to be quadratic in $V$, both equations (\ref{vinen1}) and (\ref{vinen2}) admit a solution of the form}
\be
L_T=\frac{2\Omega}{\kappa}+\alpha\left(\frac{\mathcal{B} V}{\kappa}\right)^2+O(V^3) \, ,
\ee
where $\alpha$ depends on the phenomenological parameters of the model, but is generally found to be of order unity in isotropic turbulence experiments \citep{BarenghiRev}, so we shall assume $\alpha\approx 1$, and simply consider the solution in (\ref{tot}). We will see in the following, however, that the values we will infer for the mutual friction parameter $\mathcal{B}$ would simply be rescaled as $\mathcal{B}\longrightarrow \alpha^{1/3} \mathcal{B}$, so that variations in the exact value of $\alpha$ will only weakly affect our results and will not impact on our qualitative conclusions.

In a pinned neutron star superfluid we thus expect a turbulent array of vortices, polarized by rotation, in which the vortex length per unit volume can be split into two parts $L_T=L_s+L_r$ according to (\ref{tot}). We assume that the two parts of $L_T$ can be ideally considered as a straight array of pinned vortices, with length per unit volume $L_s$, and a homogeneous and isotropic tangle with length per unit volume $L_r$ which we will approximate as formed by vortex rings, ignoring in the current analysis higher order contributions to the shape of the turbulent tangle, which is likely to exhibit a high degree of topological complexity \citep{Knots}.
Note also that different forms may be possible for the evolution equations in (\ref{vinen1}) or (\ref{vinen2}), which may lead to a difference dependence on relative velocity in the mutual friction \citep{SideryTURB, Mongiovi, Celora}, and may thus be constrained observationally, as we shall see in the following.

\section{Mutual Friction in a pinned turbulent superfluid}

Let us now turn our attention back to the mutual friction force. We will assume that on a large scale both fluids are rotating around a common axis (the axis of rotation of the star), so that the equations of motion can be written in terms of two angular velocities, $\Omega_\n$ for the neutrons and $\Omega_\p$ for the `normal' proton-electron component. {If in (\ref{totalMF}) we consider the solution $L_s$ of (\ref{rot1}) for a straight vortex array, we obtain}:
\be
F_i^\x=2\Omega_\n {\rho_\n} \mathcal{B}^{'}\epsilon_{ijk}\hat{\kappa}^j w_{\x\y}^{k}+2\Omega_\n {\rho_\n} \mathcal{B}\epsilon _{ijk}\hat{\kappa}^j\epsilon^{klm}\hat{\kappa}_l  w_{m}^{\x\y} \, 
\label{standardMF} \, ,
\ee
where, given that the vortex array is straight and the flow laminar, we have that $\hat{\kappa}=\hat{\Omega}$. This is the standard form of the anisotropic mutual friction which has been used in a number of glitch models (see e.g. \citealt{HPS12, Antonelli17, Graber18}).
For a straight vortex array we can thus write:
\beq
\dot{\Omega}_\mathrm{c}&=&2\Omega_\mathrm{n}\gamma \tilde{\mathcal{B}}(\Omega_\mathrm{n}-\Omega_\p)\frac{I_\mathrm{n}}{I_\p}-\tilde{T}\\
\dot{\Omega}_\mathrm{n}&=&-2\Omega_\mathrm{n}\gamma \tilde{\mathcal{B}}(\Omega_\mathrm{n}-\Omega_\p)+\frac{\varepsilon_\mathrm{n}}{(1-\varepsilon_{\mathrm{n}})}\tilde{T}\, ,
\label{rotatopm}
\eeq
with $\tilde{\mathcal{B}}=\mathcal{B}/(1-\varepsilon_\mathrm{n}-\varepsilon_\p)$, $\tilde{T}$ the externally induced spindown (e.g. due to the electromagnetic torque acting on the star) and $\gamma$ the fraction of free vortices with respect to the total number \citep[see e.g. ][]{HA, Khomenko}. Clearly if all vortices are pinned $\gamma=0$ and the mutual friction does not couple the fluids, leading to a lag that increases as:
\be
\frac{d}{dt} (\Omega_\mathrm{n}-\Omega_\mathrm{c})=\frac{1}{(1-\varepsilon_{\mathrm{n}})}\tilde{T}\, .
\ee
Note that in the crust of the star the entrainment parameter $\varepsilon_\n$ is likely to be large and negative, of the order of $\varepsilon_\n\approx -10$ \citep{c12} and to contribute significantly to the spin evolution \citep{aghe12, c13}, unless the crust is in a disordered state \citep{ChamelNew}.
As the lag increases, a number of instabilities can occur and are likely to destabilise the array \citep{Sidery08, Khomenkoinst},  leading to a polarized turbulent array which can be described by the solution $L_T$ in (\ref{tot}). As this solution is the sum of two parts, the polarized term $L_s$ in (\ref{rot1}) and the `isotropic' term $L_r$ in (\ref{rot2}), we assume that each segment of vortex length can be described as either polarised (s), or `ring-like` and isotropic (r) ( see e.g. \citealt{BLB12}) so that the array can be ideally seen as a combination of straight vortices and rings. 
The mutual friction in the fluid element can be split in the sum of the average over $L_s$ and the average over $L_r$, so that we have
\be
F_i^\x=L_s\left< f_i^\x\right>^{\mathrm{S}}+L_r\left< f_i^\x\right>^{\mathrm{R}}\, ,
\ee
where the superscript $S$ or $R$ indicates that the average is taken by integrating over the straight vortices $s$ or the isotropic `rings' $r$. 

Let us begin our analysis from the straight vortices case. In the limit of $\Delta\Omega\longrightarrow 0$, the {general} solution $L_T$ approaches the straight vortex array solution $L_s$ so that $L_T\approx L_s\approx 2\Omega/\kappa$ (note that if $\Delta\Omega=0$, $\Omega_\n=\Omega_\p=\Omega$). {As the lag $\Delta\Omega$ deviates from zero, and if straight vortices were originally pinned, we can expect that the straight sections will remain pinned} as long as $\Delta\Omega$ does not become large enough for the Magnus force to cause unpinning. In analogy with (\ref{standardMF}) we can thus write
\beq
F_i^\x & = & \gamma L_s \rho_\n\left[ \mathcal{B}^{'}\epsilon_{ijk}{\kappa}^j w_{\x\y}^{k} + \mathcal{B}\epsilon _{ijk}{\kappa}^j\epsilon^{klm}\hat{\kappa}_l  w_{m}^{\x\y}\right] +\nonumber\\ && + L_r\left<f_i^\x\right>^{\mathrm{R}}\nonumber\\
 & \approx& \gamma 2\Omega_\n \rho_\n\left[ \mathcal{B}^{'}\epsilon_{ijk}\hat{\kappa}^j w_{\x\y}^{k} + \mathcal{B}\epsilon _{ijk}\hat{\kappa}^j\epsilon^{klm}\hat{\kappa}_l  w_{m}^{\x\y}\right] +\nonumber\\ && + L_r\left<f_i^\x\right>^{\mathrm{R}}\, ,
 \label{mf01}
\eeq
where $\gamma$ is again the fraction of free vortex length, such that if all of the straight vorticity is pinned, and $\gamma=0$, this part does not contribute to the mutual friction, which will only be due to the average over $L_r$.

To examine this scenario in detail we recast the mutual friction per unit length of vortex line (\ref{MF}) in the form
\be
\frac{f_{i}^{\x}}{\rho_\n}= \mathcal{B}^{'}\epsilon_{ijk}\kappa^j (v_{\n}^{k}-v_\p^k)+\mathcal{B}\hat{\kappa}_i\kappa^m  (v^\n_m-v^\p_m)-\mathcal{B} \kappa (v_i^\n-v_i^\p) \, ,
\label{lineS}
\ee
where now, as the vortex can bend, we must also account for the velocity induced by its circulation. The neutron velocity can be separated in two components
\be
v^i_\n=v^i_{BK,\n}+v^i_{SI,\n}\, ,
\ee
where $v^i_{BK}$ is the background contribution due to all distant vortices, which we assume to be uniform on the scale we are considering, and $v^i_{SI}$ is the contribution due to the irrotational flow around the vortex. In the local induction approximation we only consider the contributions of nearby vortex segments, so that $v^i_{SI}$ reads
\beq
  v_i^{SI,\n}=\tilde{\tau}\epsilon_{ijk}\hat{\kappa}^{j}\hat{\kappa}^{l}\nabla_{l}\hat{\kappa}_{k}\, ,   \label{indotto}
\eeq
with  $\tilde{\tau}=\frac{1-\varepsilon_{\rm c}}{1-\varepsilon_{\rm c}-\varepsilon_{\rm n}}\mathcal{T}$, and $\mathcal{T}=\frac{\kappa}{4\pi}\log{(b/a)}$ is related to the vortex self energy, with $b$ an upper cutoff, which we take to be the intervortex spacing (of order $10^{-2}$ cm for a standard pulsar) and $a$ the lower cutoff, which we take to be the size of a vortex core, $a\approx 100$ fm. Entrainment also induces a similar term in the `normal' proton-electron fluid
\be
 v_i^{SI,\mathrm{c}}=-\frac{\varepsilon_\p}{1-\varepsilon_\p}\tilde{\tau}\epsilon_{ijk}\hat{\kappa}^{j}\hat{\kappa}^{l}\nabla_{l}\hat{\kappa}_{k}\label{indotto2} \;\text{.}
\ee

\begin{figure*}
\centerline{
 \includegraphics[width=0.49\textwidth]{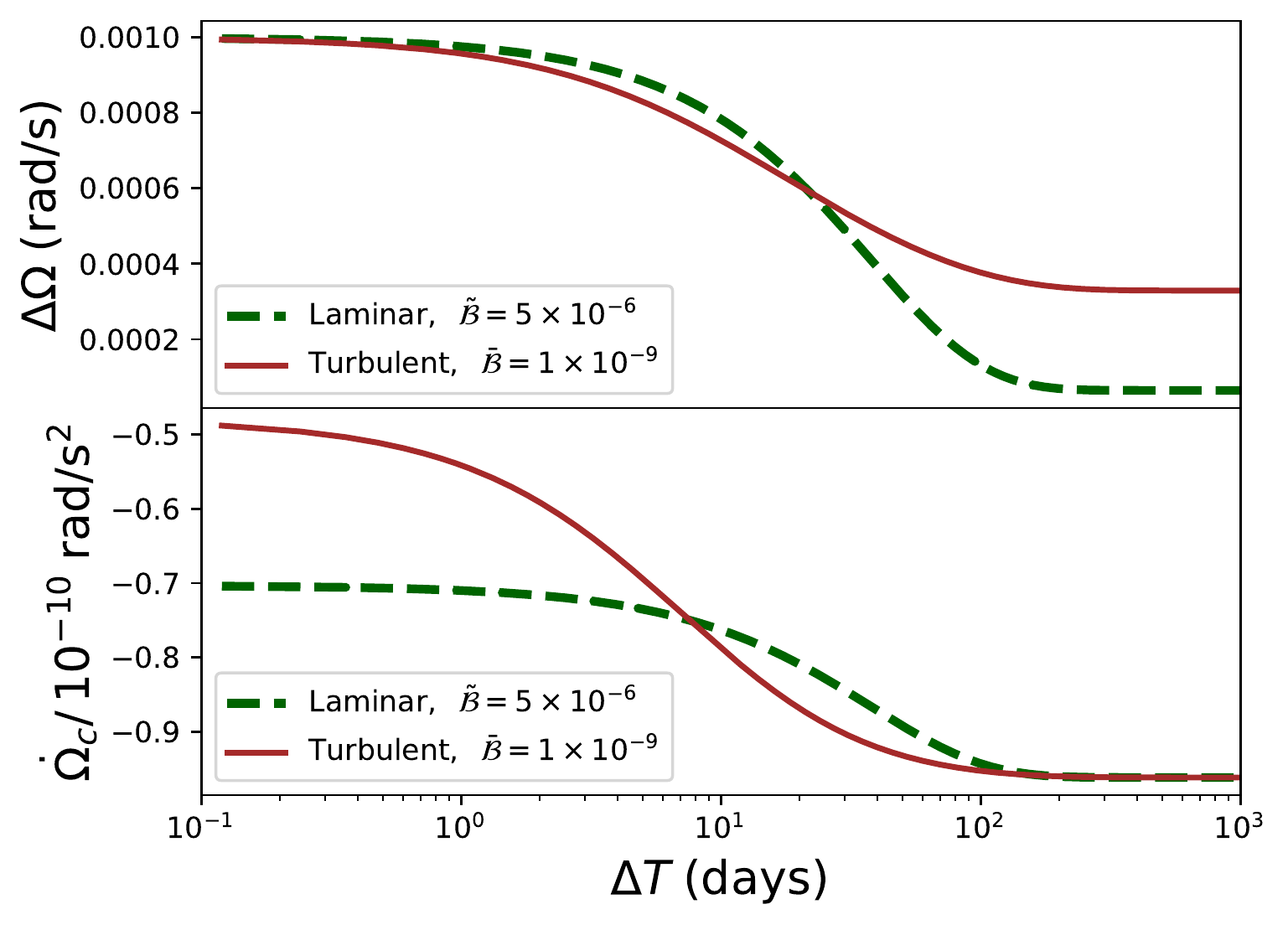} \includegraphics[width=0.49\textwidth]{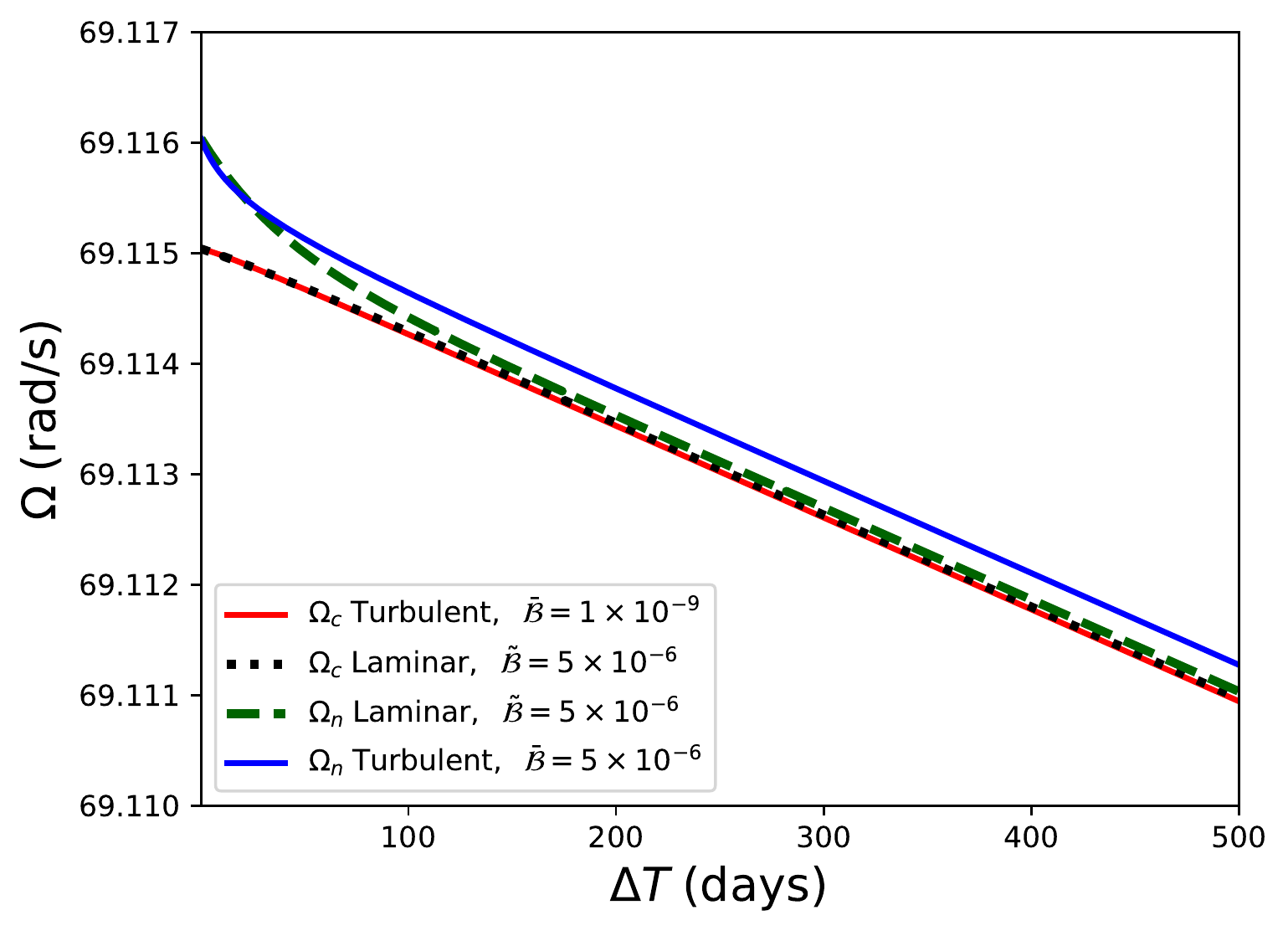}}
 \caption{Different behaviour in the laminar and turbulent model of $\Omega_\p$ and $\Omega_\n$ (right panel) and of $\Delta\Omega$ and $\dot{\Omega}_\p$ (left panel) for a strong pinning model. We see that an initial lag $\Delta\Omega_0=0.001$ rad/s rapidly disappears as the two components recouple.}
   \label{db1}
\end{figure*}

\begin{figure*}
\centerline{
 \includegraphics[width=0.49\textwidth]{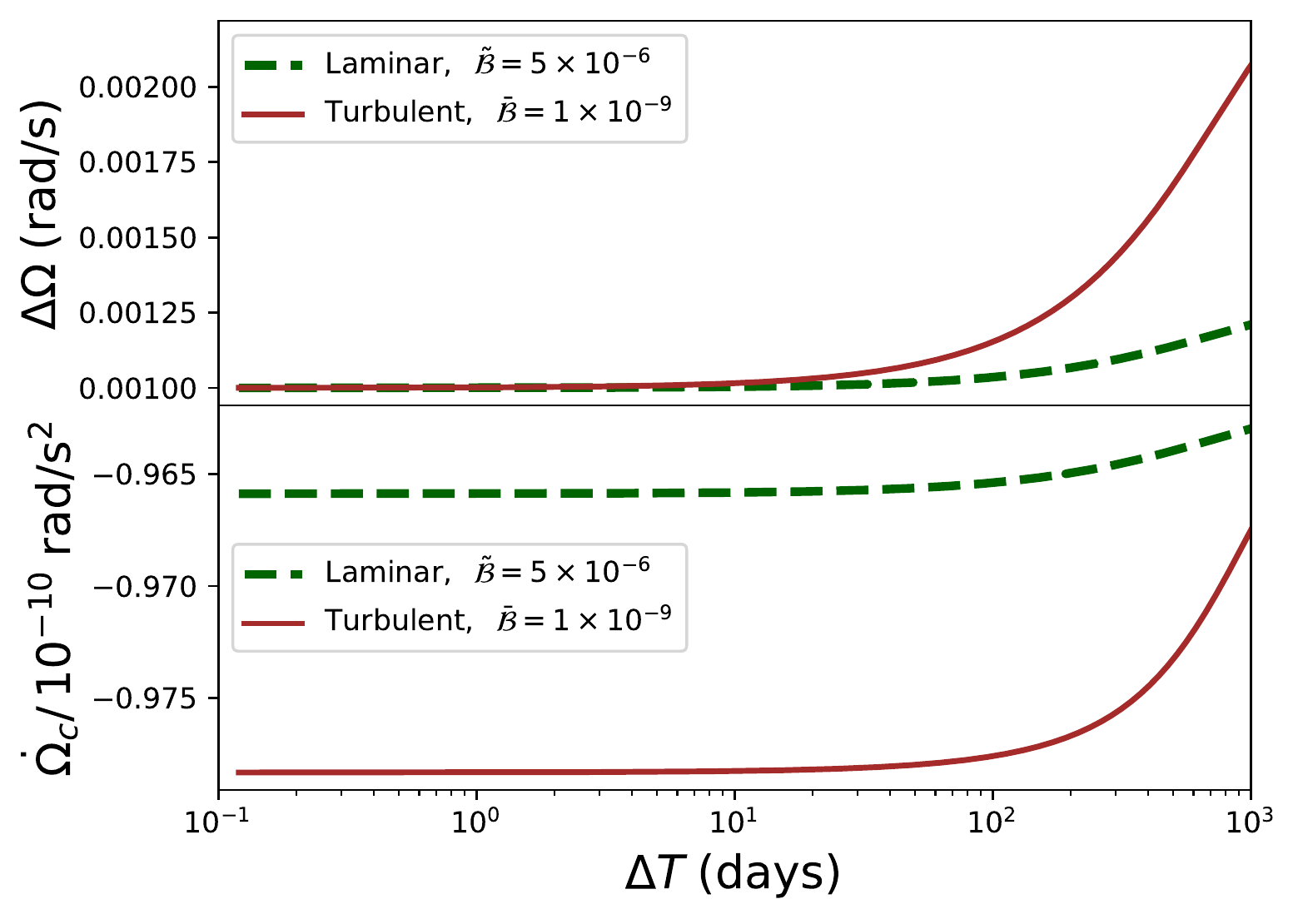} \includegraphics[width=0.49\textwidth]{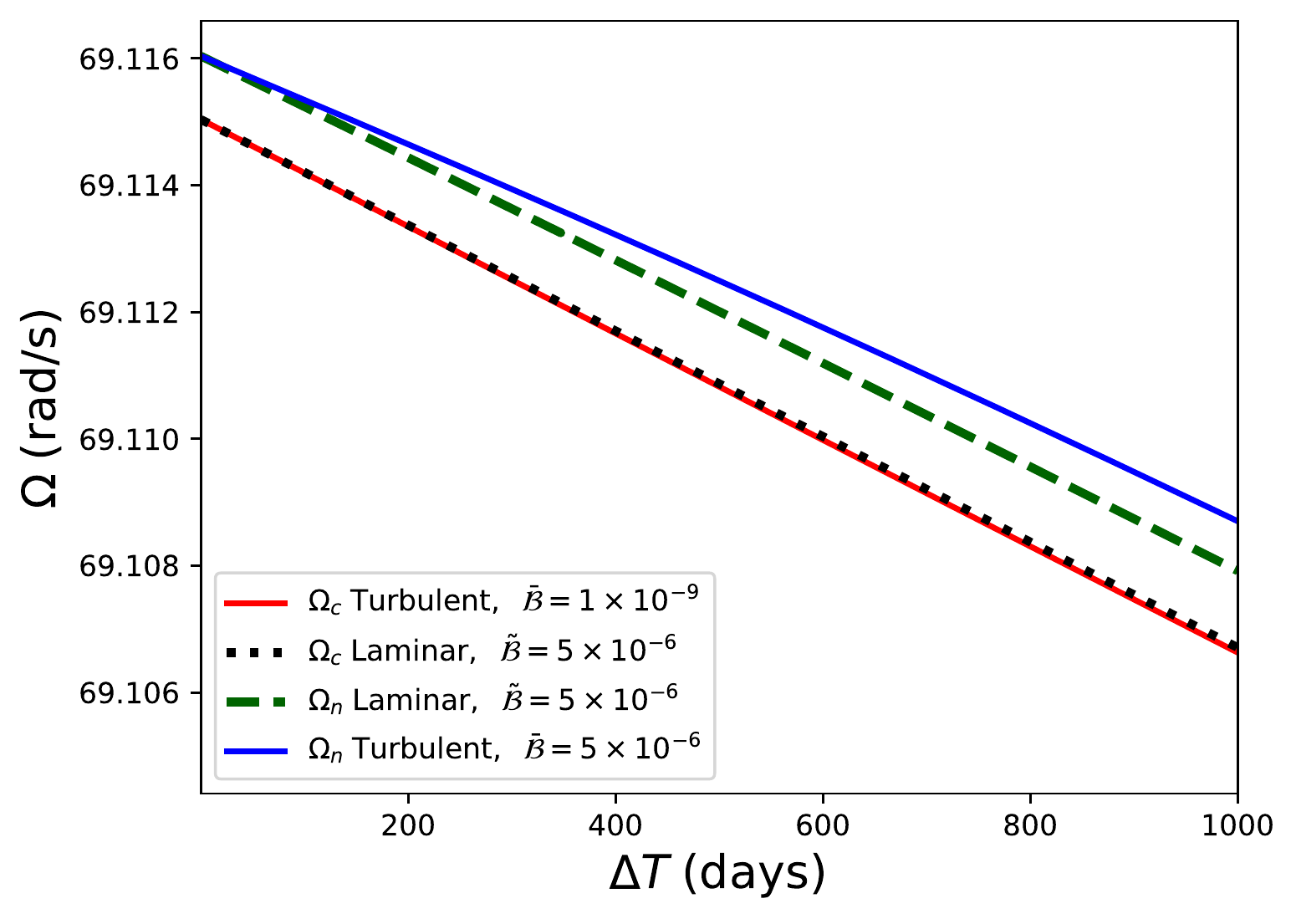}}
 \caption{Different behaviour in the laminar and turbulent model of $\Omega_\p$ and $\Omega_\n$ (right panel) and of $\Delta\Omega$ and $\dot{\Omega}_\p$ (left panel) for a weak pinning model. An initial lag $\Delta\Omega_0=0.001$ rad/s increases and the growth rate then slows down as the two components recouple.}
   \label{db2}
\end{figure*}

The effect of these terms, which we will refer to as `tension' terms, can be important in the presence of oscillations, such as Kelvin waves, in the pinned vortices (see. e.g. \citet{Jou11}). In our simplified description, however, we consider only straight vortices and rings. For the straight vortices clearly $\left<v_i^{SI,\x}\right>^{\mathrm{S}}=0$. {However, for vortex rings with random isotropic orientations, we can also assume} that averaging over the volume one has $\left<v_i^{SI,\x}\right>^{\mathrm{R}}\approx 0$. {We thus neglect the contributions of the self induced velocity in the following, and together with the fact that isotropy also implies} $\left<\hat{k}_i \right>^{\mathrm{R}}\approx 0$, we have that
\be
\left< \frac{f_{i}^{\x}}{\rho_\n} \right>^{\mathrm{R}}\approx -\mathcal{B} \kappa (v_i^{BK,\n}-v_i^{BK,\p}) \;\text{.}
\ee
Hereafter we will drop the superscript `${BK}$'. The mutual friction force per unit volume follows from (\ref{mf01}) 
\beq
F_i^\x & = & \gamma L_s \rho_\n\left[ \mathcal{B}^{'}\epsilon_{ijk}{\kappa}^j w_{\x\y}^{k} + \mathcal{B}\epsilon _{ijk}{\kappa}^j\epsilon^{klm}\hat{\kappa}_l  w_{m}^{\x\y}\right] +\nonumber\\ && - L_r\rho_\n \mathcal{B} \kappa w_i^{\x\y}\\\
& \approx& \gamma 2\Omega_\n \rho_\n\left[ \mathcal{B}^{'}\epsilon_{ijk}\hat{\kappa}^j w_{\x\y}^{k} + \mathcal{B}\epsilon _{ijk}\hat{\kappa}^j\epsilon^{klm}\hat{\kappa}_l  w_{m}^{\x\y}\right] +\nonumber\\ && - {\alpha} \rho_\n \frac{\mathcal{B}^3}{\kappa} (w_j^{\x\y} w^j_{\x\y}) w_i^{\x\y}\, .
\eeq
Note that the factor $\gamma$ only applies to {the straight vortices, and not to the length of vortex rings in our fluid element, which thus }always contribute to the mutual friction. If we again assume that both fluids are rotating around a common axis, and take $V\approx R\Delta\Omega$, where $R$ is taken to be approximately the stellar radius, we have
\beq
\dot{\Omega}_\mathrm{c}&=&2\Omega_\mathrm{n}\gamma {\tilde{\mathcal{B}}}\Delta\Omega\frac{I_\mathrm{n}}{I_\mathrm{c}}+{\alpha}\frac{R^2}{\kappa} \bar{\mathcal{B}}^3\Delta\Omega^3\frac{I_\mathrm{n}}{I_\mathrm{c}}-\tilde{T}\\
\dot{\Omega}_\mathrm{n}&=&-2\Omega_\mathrm{n}\gamma {\tilde{\mathcal{B}}}\Delta\Omega-{\alpha}\frac{R^2}{\kappa}{\bar{\mathcal{B}}^3}\Delta\Omega^3+\frac{\varepsilon_\mathrm{n}}{(1-\varepsilon_{\mathrm{n}})}\tilde{T}\, ,
\label{main}
\eeq
where $\tilde{\mathcal{B}}=\mathcal{B}/(1-\varepsilon_\n-\varepsilon_p)$ and $\bar{\mathcal{B}}=\mathcal{B}/(1-\varepsilon_\n-\varepsilon_p)^{1/3}$. This is our main result. We see from (\ref{main}) that there is always a dissipative contribution from the turbulent array (equivalent to the standard Gorter-Mellink form of the mutual friction for isotropic turbulence), even when the superfluid is pinned on the mesoscopic scale. In fact, whilst this contribution is generally weaker that the anisotropic contribution due to the $\Delta\Omega^3$ dependence, it becomes dominant in the perfect pinning regime when $\gamma\approx 0$.
In figures (\ref{db1}) and (\ref{db2}) we show the evolution of the system, given an initial lag $\Delta\Omega_0$, for the case of strong and weak mutual friction respectively, for both the laminar ($\alpha=0$) and pinned turbulent ($\gamma=0$) case. We can see that the evolution of $\Omega_\n$ and of the lag $\Delta\Omega$ varies between the turbulent and laminar model, but these variables cannot be observed directly. The evolution of $\Omega_\p$, which we identify with the observable component of the pulsar spin, on the other hand does not vary significantly between the two models. The evolution of the derivative $\dot{\Omega}_\p$, however, varies and if identified with the observable frequency derivative of the neutron star, can be probed by pulsar timing, as we will discuss in the following. 

\section{Pulsar glitch recoveries}

In this section we consider a practical application of the theory developed above to the post-glitch recovery of pulsars. Previous studies \citep{HPS12,Graber18, Montoli19} have focused on the standard form of mutual friction for straight vortices, which give an exponential response to the glitch perturbation, as we will see below. In this case, the post-glitch evolution of the spin and spin-down rate can be described as a sum of exponentials. However, in some pulsars -- and most typically for large glitches like those of the Vela pulsar -- the post-glitch response is better characterised by an initial exponential relaxation followed by what appears to be a linear decrease in the magnitude of the spin-down rate. This behaviour is hard to model in terms of the straight vortex mutual friction. Moreover, it has been suggested that the large glitches in the Vela pulsar originate in the core, not the crust, of the star \citep{aghe12, gug14, nbh15, crab}. Here, we suggest that if indeed the glitch originates in an interior (possibly core) region, the resulting decrease in lag in the crust will stop the flow of free vortices, and this region will thus not respond exponentially. A similar scenario of vortex flow cessation was explored by \citet{nonlinear}, who suggested that  non-linear recoupling due to vortex creep may explain the linear recovery. We take a different view and propose that it is the non-linear response due to the {\it turbulent} array that drives the quasi-linear relaxation.

First, let us consider a region of superfluid with moment of inertia $I_\n$, coupled to a normal component with moment of inertia $I_\p$ as in equations (\ref{rotatopm}); that is, for straight vortex mutual friction with $\alpha=0$. To simplify the analysis we consider only the short term dynamics and neglect the external spin-down term $\tilde{T}$. The evolution equation for the lag $\Delta\Omega$ takes the form
\be
\Delta\dot{\Omega}=-2\Omega_\n\gamma\frac{\mathcal{B}}{(1-\varepsilon_\n-\varepsilon_\p)}\Delta\Omega\frac{I_\p+I_\n}{I_\p}\, ,
\ee
which, taking a constant value for $\Omega_\n$ (i.e. neglecting the small change in frequency due to the glitch compared to the stellar frequency), has exponential solutions of the form
\be
\Delta\Omega=\Delta\Omega_0 e^{-t/\tau}\, ,
\ee
where $\Delta\Omega_0$ is the initial lag, due to the glitch, and 
\be
\tau=\frac{I_\p (1-\varepsilon_\n-\varepsilon_\p)}{2\Omega_\n\gamma{\mathcal{B}}(I_\n+I_\p)}\, .
\label{taus}
\ee
The region of superfluid with moment of inertia $I_\n$ thus re-couples exponentially to the bulk of the star that is spinning down. 
If we identify the observed spin rate of the pulsar with the spin rate of the `normal' ($\p$) component, to which the magnetic field and radio emission are tied, we may approximate the spin-down rate as:
\be
\dot{\Omega}=\dot{\Omega}_{\p}=-\frac{T_{sd}}{I_T(1+\frac{I_\n}{I_T}(1-\exp{(-t/\tau)}))}\, ,
\label{sdapp}
\ee
where $T_{sd}$ is the external spin-down torque, which acts only on the part of the total moment of inertia that is coupled to the crust at a given time. Before a glitch, the region with moment of inertia $I_\n$ is coupled to the rest of the star (which we assume to have moment of inertia $I_T$ - this includes all the normal component, but also superfluid components coupled on short timescales in the core), and they spindown together at a rate $\dot{\Omega}=-T_{sd}/(I_T+I_\n)$. When a glitch is triggered in the core, the spin frequency of the `normal' component of the crust rises, and the region $I_\n$ decouples.  
If $t/\tau << 1$, then (\ref{sdapp}) can be approximated as a linear evolution of the spin-down rate
\be
\dot{\Omega}\approx -\frac{T_{sd}}{I_T}\left(1-\frac{{I}_{\n}}{I_T} \frac{t}{\tau}\right)\, ,\label{sdapp2}
\ee
and the observed braking index $n=\ddot{\Omega}\Omega/\dot{\Omega}^2$, will be
\be
n\approx \frac{I_\n}{I_T}\frac{\Omega}{\tau |\dot{\Omega}|}\, ,
\label{braking_straight}
\ee
where in the last step we have assumed $(I_n/I_T)(t/\tau)<<1$. 
For the Vela pulsar, typical values for the observable parameters are $n\sim40$ (between glitches) \citep{ES17}, $\dot\Omega/(2\pi)=\dot\nu=-1.56\times 10^{-11}$ s$^{-2}$ and $\Omega/(2\pi)=\nu=11$ s$^{-1}$, where $\nu$ and $\dot{\nu}$ are the rotational parameters of the pulsar obtained by timing the neutron star's pulsations from radio observations. The inter-glitch time interval is usually of order $\sim 3$ years thus for our approximation $t/\tau<<1$ we require $\tau\gtrsim 10^8$ s. Therefore, from (\ref{braking_straight}) we find $I_\n/I_T\gtrsim 0.005$, which would be consistent with a fraction of the crust giving rise to the linear relaxation of $\dot{\nu}$. 
However, for this to hold, the condition $\tau\gtrsim 10^8$ implies that
\be
\frac{I_\p}{(I_\n+I_\p)}\frac{(1-\varepsilon_\n-\varepsilon_\p)}{2\Omega_\n\gamma\mathcal{B}} \gtrsim 10^8\, ,
\ee
which leads to 
\be
\frac{\gamma\mathcal{B}}{(1-\varepsilon_\n-\varepsilon_\p)}\lesssim 7\times 10^{-11}\, .
\ee
These values are in tension with theoretical expectations. In the crust one expects to have $\varepsilon_\n \approx -4$ (which leads to $\varepsilon_\p=\varepsilon_\n I_\n/I_\p\approx 0$ for  $I_\n/I_\p\approx 0.01$)  \citep{c12, Khomenkoinst}, and  in the presence of large lags due to pinning, the mutual friction is mainly due to kelvon excitations, giving $\mathcal{B}\approx 10^{-3}$  \citep{JonesKelvon, EBKelvon, Graber18}. For excitations of the lattice one has $10^{-9}\lesssim\mathcal{B}\lesssim 10^{-5}$  \citep{Jones90, Jones91}, although this mechanism requires low relative velocities of the free vortices, which are not expected if large lags develop due to pinning and the kelvon excitations described above become the main dissipative channel. Nevertheless, even if phonon excitations are the dominant mechanism, one would need a very low fraction of free vortices $\gamma$ to be compatible with our observational estimate, and the turbulent terms will become sizable in this situation.  We thus expect a clear exponential response from regions with a straight vortex array. If pinning occurs in the outer core the required value of $\gamma$ is even smaller, as in this case $\varepsilon_\n\approx\varepsilon_\p\approx 0$ and one expects mutual friction to be due mainly to kelvon excitations ($\mathcal{B}\approx 10^{-3}$) or electron scattering off vortex cores ($\mathcal{B}\approx 10^{-4}$) \citep{HasSed}. This scenario is thus not favoured by our theoretical understanding of the strong pinning regions in a neutron star crust or core.

The situation is different in the presence of polarized turbulence. In this case the equations of motion for the lag between two different fluids, on short timescales on which we neglect the effect of the external spin-down, take the form:
\be
\Delta\dot{\Omega}=-{\alpha}\frac{R^2}{\kappa}\frac{\mathcal{B}^3}{(1-\varepsilon_\n-\varepsilon_\p)}\Delta\Omega^3\frac{I_\p+I_\n}{I_\p}\, ,
\ee
which has solutions of the form
\be
\Delta\Omega=\frac{\Delta\Omega_0}{\sqrt{1+t/\tau_t}}\, ,
\ee
where $\Delta\Omega_0$ is the initial lag, and 
\be
\tau_t=\frac{\kappa (1-\varepsilon_\n-\varepsilon_\p)}{{\alpha}R^2 \mathcal{B}^3}\frac{I_\p}{ (I_\n+I_\p)} \frac{1}{\Delta\Omega^2_0}\, .
\label{tauT}
\ee
We can write a similar expression to (\ref{sdapp}) for the spindown:
\be
\dot{\Omega}_{\p}=-\frac{T_{sd}}{I_T\left(1+\frac{I_\n}{I_T}\left(1-1/\sqrt{1+t/\tau_t}\right)\right)}\label{sdappT}\, ,
\ee
and for $t/\tau_t << 1$, we obtain a similar form as (\ref{sdapp2})
\be
\dot{\Omega}\approx -\frac{T_{sd}}{I_T}\left(1-\frac{{I}_{\n}}{2 I_T} \frac{t}{\tau_t}\right)\, ,
\label{sdapp2T}
\ee
and can repeat the same analysis. Now however, the condition $t/\tau_t << 1$, which guarantees a linear evolution in $\dot{\nu}=\dot{\Omega}/(2\pi)$ associated with a large value of the braking index $n$, reads (for $t=10^8$ s as before, and taking as a typical value $\Delta\Omega_0=10^{-4}$)
\be
\frac{{\alpha}\mathcal{B}^3}{(1-\varepsilon_\n-\varepsilon_\p)}\lesssim 2\times 10^{-7}\,
\ee
which is easily compatible with conditions in the crust \citep{Graber18}, assuming $\alpha\approx 1$ ($\Rightarrow\;\mathcal{B}\lesssim 10^{-2}$)  \citep{SideryTURB, Sciacca08}.

We can also see, from equations (\ref{sdapp}) and (\ref{sdapp2T}) for each model, that in either case the second derivative of the spin frequency scales as $\ddot{\nu}\propto {\tau^{-1}}$
with $\tau$ the respective timescale for straight vortices or turbulence. In the straight vortex case $\tau$ is independent of the initial lag between the fluids  (Eq. \ref{taus}) but in the turbulent case it depends inversely on the square of the initial lag $\Delta\Omega_0$ (Eq. \ref{tauT}), so at late times following a glitch $\ddot{\nu}$ depends approximately quadratically on $\Delta\Omega_0$,
\be
\ddot{\nu}\propto \Delta\Omega_0^2\, .
\label{propo2}
\ee 
The quantity $\Delta\Omega_0$ is not experimentally accessible, as the rotation rate of the neutron condensate is not directly observable. Here, we attempt to approximate it by considering the lag that would have been built up since the previous glitch (assuming perfect pinning), $\Delta\Omega_0\approx\Delta T \dot{\Omega}_\p /(1-\varepsilon_\n)$, with $\Delta T$ the waiting time since the previous glitch (note that we are neglecting both the effect of the increased spin-down after a glitch, and the change on frequency due to the glitch itself, which provide only a small correction).
If we are always observing the response of (roughly) the same region of the star, we thus expect to find a correlation between the measured value of $\ddot{\nu}$ in the late, `linear' phase of the evolution after a glitch, and the waiting time from the previous glitch. 
Note that this approximation is likely to be crude, as not only does it neglect the effect of the glitch itself and of the variations in spindown rate, but also, by just considering the lag built up since the previous glitch, the cumulative effect of subsequent glitches. These effects are likely to be important, especially for longer datasets, and will be considered in future work. Nevertheless our model captures the main features of the turbulent response of the superfluid, and predicts an observable effect.

\begin{figure}
\centerline{
 \includegraphics[width=0.49\textwidth]{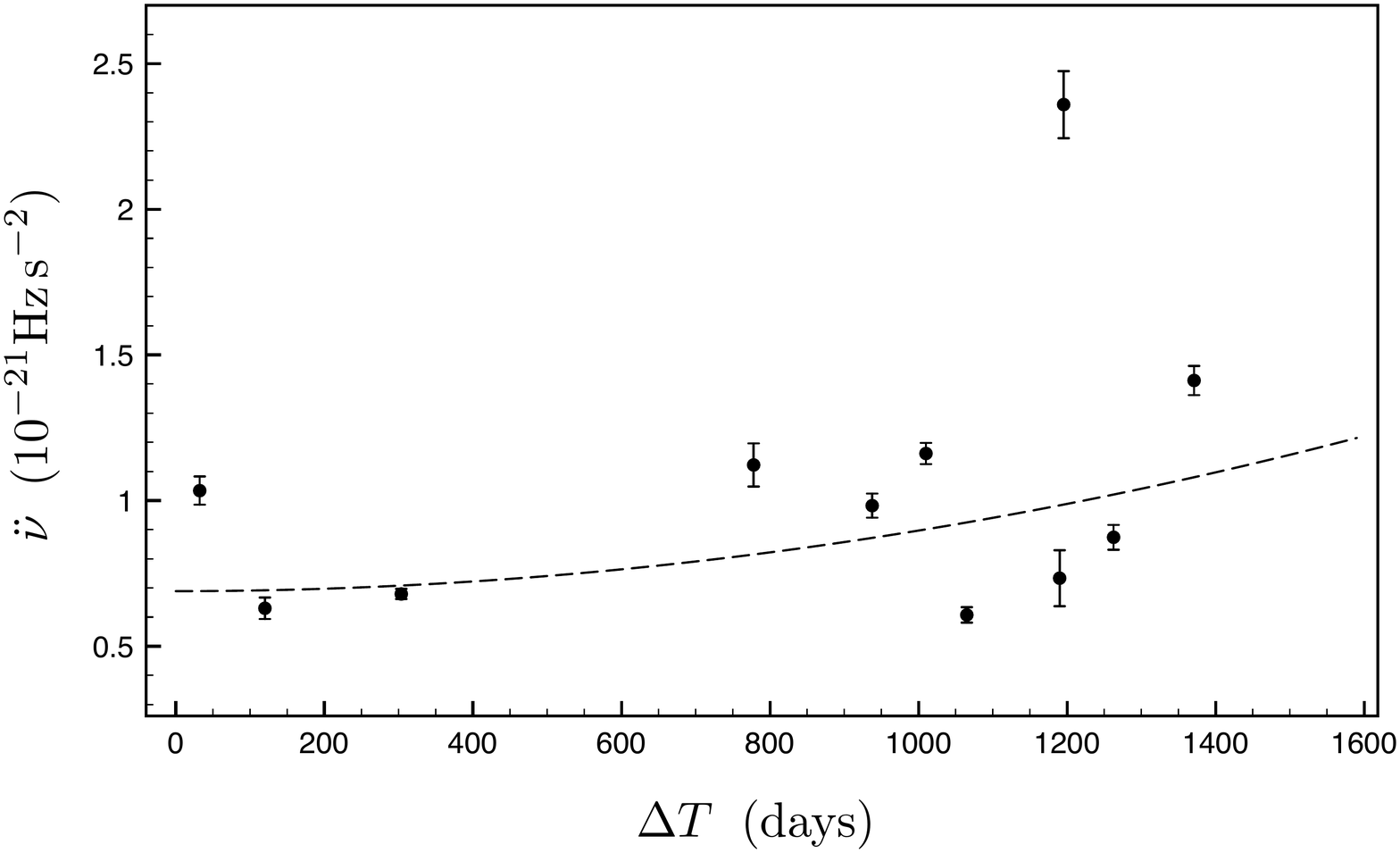}}
 \caption{Vela pulsar: Measured second frequency derivative $\ddot{\nu}$ for the quasi-linear recovery part after the Vela glitches between 1969 and 2004 \citep{ES17}, plotted versus the inter-glitch waiting time preceding each glitch. The dashed line corresponds to the best-fit quadratic function of the form $\ddot{\nu}/(10^{-21} s^{-2})=a+b (\Delta T)^2$. We obtain $a=0.689$, $b=2.08 \times 10^{-07}$  days$^{-2}$, corresponding to $\mathcal{B}\approx 2\times 10^{-6}$. Details of the data used can be found in Appendix \ref{appendix}.}
   \label{figvela}
\end{figure}

\begin{figure}
\centerline{
 \includegraphics[width=0.49\textwidth]{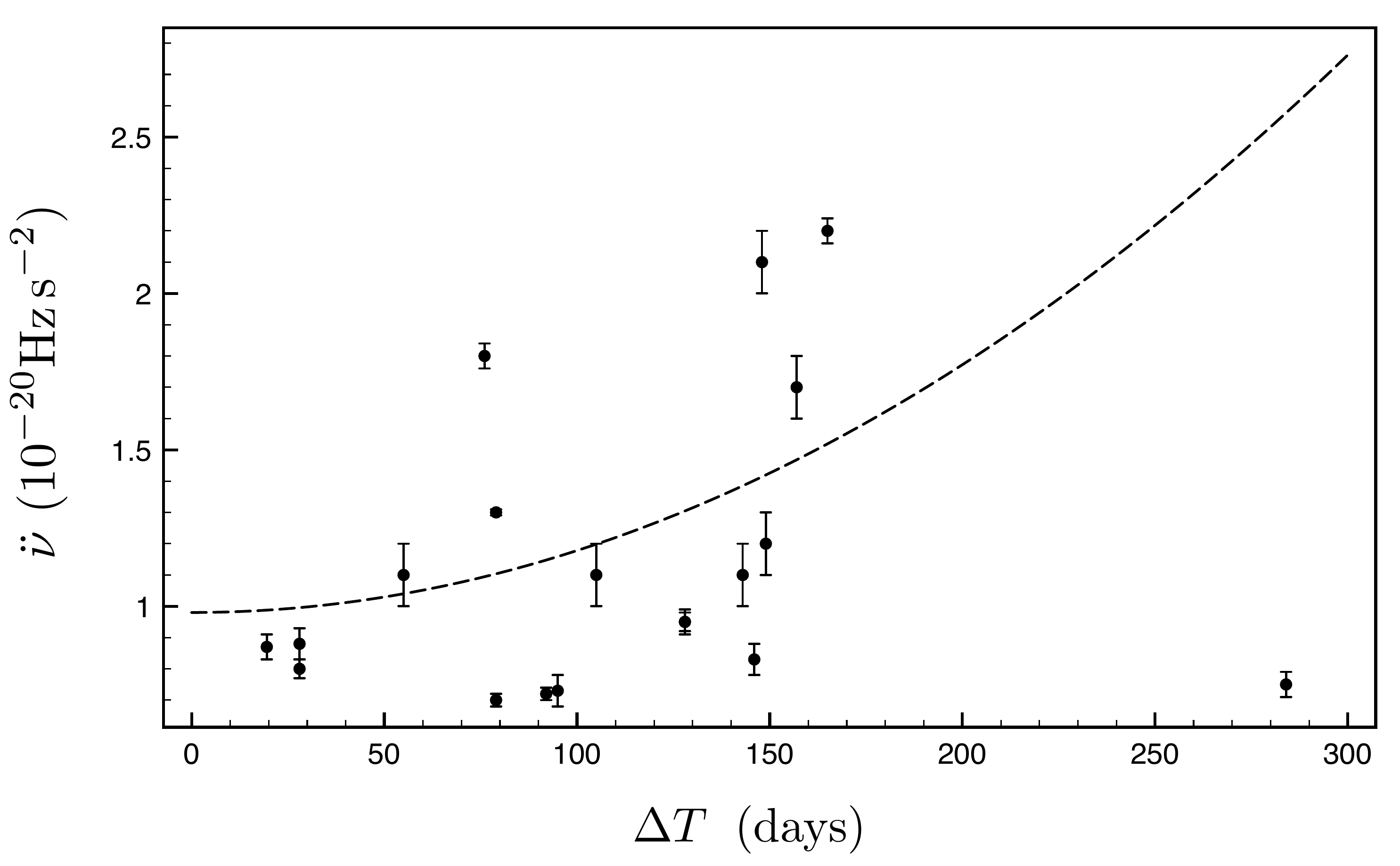}}
 \caption{PSR J0537-6910: measured frequency second derivative after a glitch from \citet{aek+18}, excluding fits to intervals with less than 10 measured TOAs and/or errors greater than 10\%, plotted versus the waiting time in days since the previous glitch. We use data from \citet{aek+18}, however recent observations by {\it NICER} have detected additional glitches \citep{NicerWynn}. The addition of these new data points does not modify our conclusions, and is presented in Appendix \ref{appendix}. 
 We plot also a fitted quadratic function of the form $\ddot{\nu}/10^{-20} s^{-2} =a+b (\Delta T)^2$, for which we obtain $a=0.957$  and $b=1.94\times 10^{-05}$ days$^{-2}$, which for standard parameters described in the text gives an estimate of $\mathcal{B}\approx 2 \times 10^{-6}$ for the strength of the mutual friction, consistent with estimates for the crust of the neutron star.}
   \label{fig0537}
\end{figure}
\begin{figure*}
\centerline{
 \includegraphics[width=0.48\textwidth]{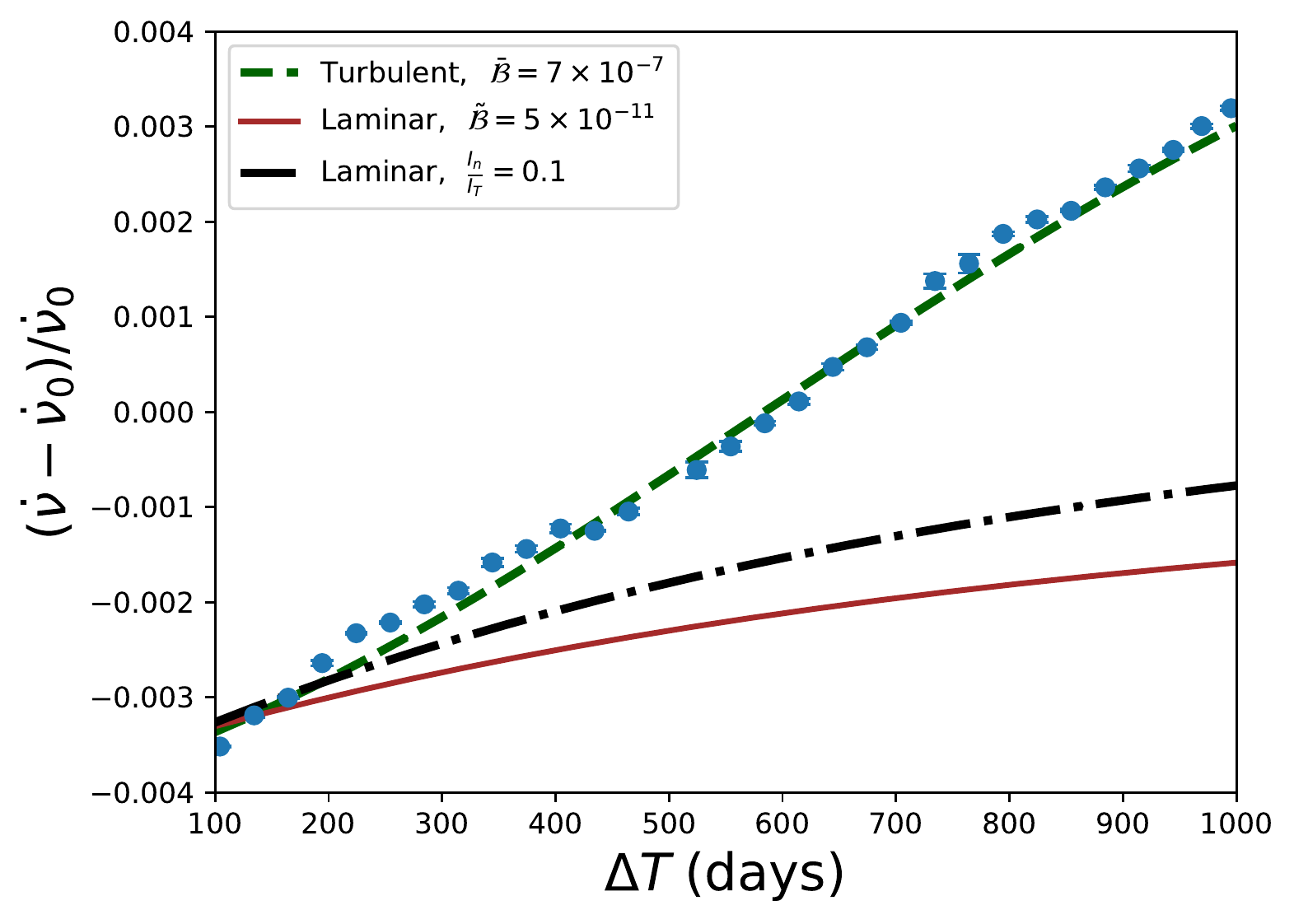} \includegraphics[width=0.48\textwidth]{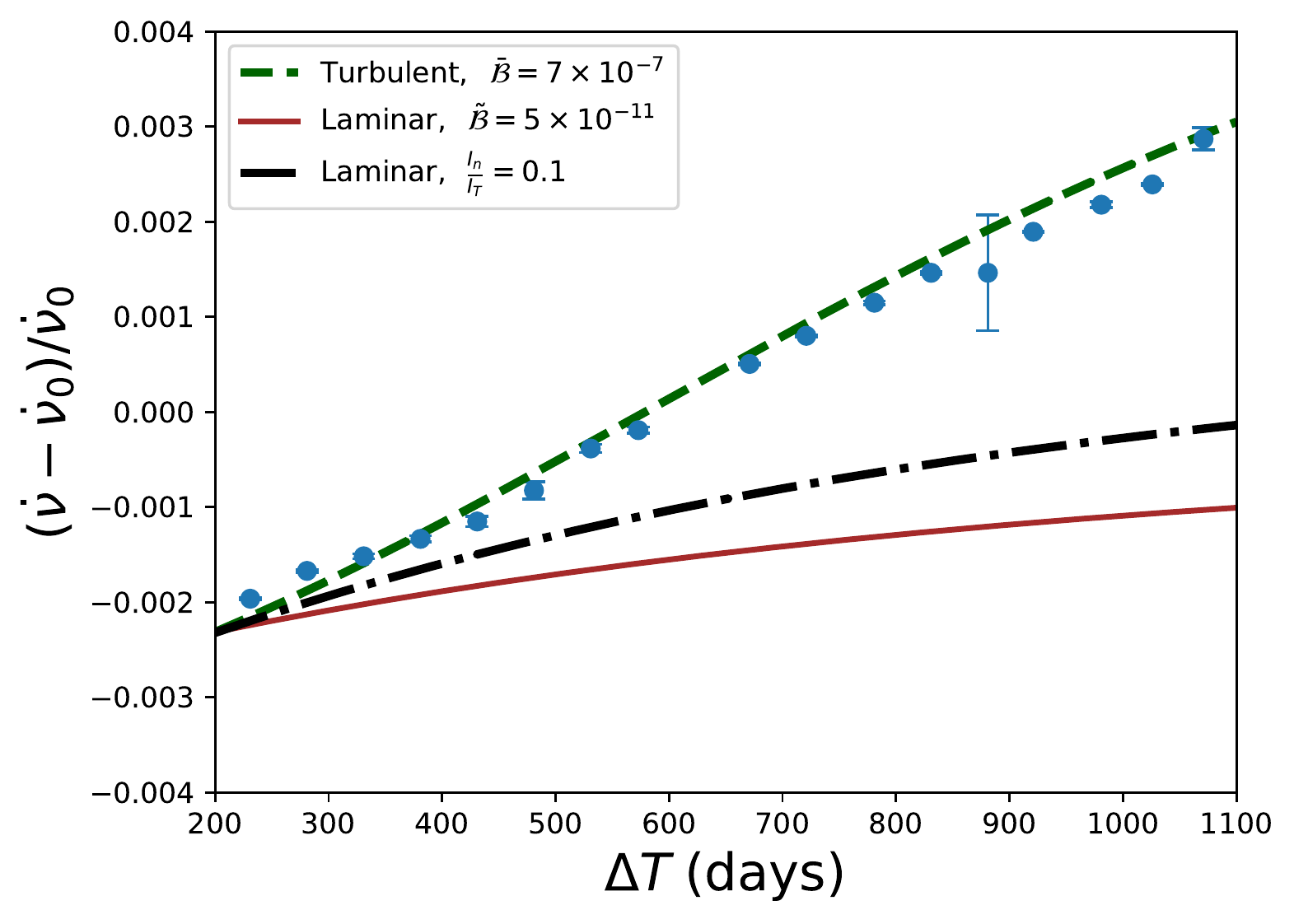}}
 \caption{Example of the evolution of $\dot{\nu}$ after the 1975 (left) and 1996 (right) glitches of the Vela pulsar, which we have selected as they display a long quasi-linear recovery. We plot both a model with straight vortices and a polarized turbulent model, as described in the text. We exclude the initial phase of the post-glitch recovery, which is quasi-exponential and most likely driven from a different region (e.g. the core). We see that for standard parameters the turbulent model is a good fit for the data, while the straight vortex laminar model is not, even if we allow for a high ratio of $I_\n/I_T=0.1$, which is unlikely to be achieved in the crust, and would correspond to part of the core decoupling. In fact, we cannot fit the observed trend with a linear model, unless we allow for unrealistically high values of $I_\n/I_T \gtrsim 0.2$.}
   \label{fig2}
\end{figure*}

In figures (\ref{figvela}) and (\ref{fig0537}) we plot the measured values of the $\ddot{\nu}$ versus waiting time for the Vela pulsar and PSR J0537-6910, which are two of the most prolific glitchers, and both of which show a `regular' glitching behaviour, with a majority of large glitches that occur quasi-periodically \citep{Howitt18}. Although no statistically robust conclusions can be drawn due to the scatter of the data, which is anyway to be expected also due to the simplifying assumptions that enter our model, we note a trend towards larger values of $\ddot{\nu}$ for longer waiting times. 

We fit a quadratic function of the form $\ddot{\nu}=a+b(\Delta T)^2$ to our data, and estimate the value of $\mathcal{B}$ from (\ref{tauT}).  Assuming $R=10$ km, $I_n=0.01 I_T$, and using the measured reference values, obtained from the ATNF pulsar catalogue \footnote{https://www.atnf.csiro.au/research/pulsars/psrcat/} \citep{ATNF} for $\dot{\nu}$ and $\nu$ (for Vela $\nu=11.19$ s$^{-1}$, $\dot\nu=-1.57\times 10^{-11}$ s$^{-2}$ and for J0537-6910 $\nu=62.03$ s$^{-1}$, $\dot\nu=-1.99\times 10^{-10}$ s$^{-2}$ ) , we obtain $\mathcal{B}\approx 2.0 \pm 0.7\, \times  10^{-6}$ for the Vela pulsar and  $\mathcal{B}\approx 1.8 \pm 0.5\, \times  10^{-6}$ for PSR J0537-6910, which are consistent with each other and with the theoretically expected values of the mutual friction in the crust \citep{HasSed}.  Pulsar J0537-6910 deserves, however, a separate discussion. From figure (\ref{fig0537}) one can see that the last data point, $\Delta T=284$ days, deviates significantly from our fit, which was, in fact, performed only on data with $\Delta T\leq 170$ days. This is because the relation in (\ref{propo2}) that we are fitting, relies on the approximation $t \ll \tau$. Given that $\tau\propto \Delta T^2$, we are, in fact, implicitly assuming that if the condition $t\ll \tau$ holds for a particular glitch, it will hold for all glitches with similar values of $\Delta T$. The last data point for J0537-6910, however, corresponds to a waiting time almost double that of the other data points, and will result in a timescale $\tau$ which is much shorter that for the other post-glitch relaxations (at least a factor of 4). In this case we cannot truncate the expansion of (\ref{sdappT}) at the linear order. Taking the expansion to second order we have:
\be
\ddot{\Omega}_\p\approx \frac{I_\n}{2 I_T}\frac{1}{\tau}-\left(\frac{3}{4}+\frac{1}{2}\frac{I_\n}{I_T}\right)\frac{t}{\tau^2}+O\left(\frac{t}{\tau}\right)^3
\ee
so that for $t\approx \tau$ the value of the second derivative is reduced with respect to the first order fit, as is indeed seen in figure (\ref{fig0537}).
We can also perform a further consistency check. The point with $\Delta T=284$ days corresponds to the second glitch observed in the pulsar, and was followed by a waiting time of 149 days until the next glitch. We can then use the condition that $t\approx \tau$, when the second order term becomes important, to evaluate independently the mutual friction parameter $\mathcal B$ from (\ref{tauT}). Using standard parameters from the previous section we obtain $\mathcal B\approx 10^{-6}$, which is compatible with our earlier estimates.

This result provides additional evidence for the picture that in both stars we are observing the same crustal region responding to a glitch which was triggered elsewhere, possibly in the core, as suggested also by \citet{aghe12, c13, crab}.
Note that for both pulsars, if one considers standard mutual friction due to straight vortices, $\ddot{\nu}$ should be independent from waiting time, and there would be no reason for $\ddot\nu$ to be correlated with $\Delta T$, as the timescale $\tau$ is a constant in this model. Furthermore, in the straight vortex case the required values of the effective mutual friction are in the range $10^{-12}\lesssim \gamma\mathcal{B}\lesssim 10^{-11}$, which are very low for standard crustal models, even assuming phonon mediated mutual friction in the presence of pinning \citep{HPS1}.

We can also easily verify our approximations by integrating numerically the equations in (\ref{main}). An example of the results is given in Figure (\ref{fig2}), where we compare three models to the evolution of $\dot\nu$ (once the initial post-glitch strong exponential evolution is over) following the 1975 and 1996 Vela glitches, which we select as they display a long quasi-linear recovery after the glitch (see Appendix  \ref{appendix} for a description of the data used). For the turbulent model we assume $\alpha=1$, $\gamma=0$ and employ only the expression for the turbulent mutual friction, setting $\varepsilon_\n=-4$ \citep{aghe12}, $I_\n/I_T=0.05$, $\bar{\mathcal{B}}=\mathcal{B}/(1-\varepsilon_\n-\varepsilon_\p)=7\times 10^{-7}$ (corresponding to $\mathcal{B}\approx 10^{-6}$, consistently with our previous estimates) and the initial lag is set from the expression $\Delta\Omega_0\approx\Delta T \dot{\Omega}_\p /(1-\varepsilon_\n)$, with $\Delta T = 1342$ days for the 1975 glitch, and $\Delta T = 819$ days for the 1996 glitch.  For the laminar model on the other hand we take $\gamma=1$, $\alpha=0$, an  as in previous estimates $\tilde{\mathcal{B}}=\mathcal{B}/(1-\varepsilon_\n-\varepsilon_\p)=5\times 10^{-11}$. We include a final laminar model in which we take the unrealistically high value $I_\n/I_T=0.1$, which would correspond to a large fraction of the core being decoupled. As can be seen from Figure (\ref{fig2}) the turbulent model is a good fit to the quasi linear evolution over a wide range of data, while the laminar case is generally not, even if we allow for $I_\n/I_T=0.1$ (although note that also in this case the exponential trend is still observable). In order for the laminar model to fit more closely the trend, one needs to decouple high fractions of the core moment of inertia, above $\approx 20\%$, which is unrealistically high for most glitch models \citep{hm15}.
In conclusion, our numerical experiments also confirm that the turbulent model provides a better fit to the late time post glitch relaxations of the Vela pulsar.

\section{Conclusions}

Neutron star crusts represent a complex system, in which superfluid neutrons flow through an array of pinning sites, that are vastly more abundant than the superfluid vortices which carry the circulation \citep{HasSed}. In such a system superfluid turbulence is expected to develop \citep{SideryTURB}, aided by the presence of pinning sites at which vortex rings may be preferentially created \citep{Stagg17}. If that is the case, then to describe the neutron star crust and how it may respond to a glitch, it is necessary to model a polarized, turbulent array of pinned vortices.
The main result of this paper is, in fact, that astrophysical observations of the late time (after the initial strong exponential response is over) post glitch relaxations of large pulsar glitches are better described in terms of a polarized and turbulent vortex array model, and not in terms of the laminar spin down of a straight vortex array.

In this paper we have developed a theory of mutual friction for a polarized turbulent superfluid, allowing for pinning throughout the volume occupied by the fluid. {Unlike in superfluid Helium experiments, where pinning occurs at the boundaries of the container, in a neutron star, pinning occurs in the bulk of the condensate due to density inhomogeneities in the normal fluid (e.g. the nuclear clusters in the crust). This introduces perturbations on the vortex at these small scales, that lead indirectly to increased dissipation via the mutual friction} To study this system we have expanded the vortex length per unit volume in two components, a straight one which contributes to the overall circulation and remains pinned, and an additional isotropic component of vortex `rings', which do not pin, but contribute to the mutual friction. This division is, of course, idealised, and neglects higher order contributions \citep{Knots}, which however are expected to be small for the small difference in velocity between the superfluid and the normal component that exists in a neutron star.
{In summary, we should think of pinning as the continual injection of high curvature. In our model, the injection of vortex rings.} 

We find that, in the presence of strong pinning, the mutual friction takes the isotropic Gorter-Mellink form, which does not lead to exponential recoveries as in the standard laminar case considered in most glitch cases, but rather a power-law recovery \citep{SideryTURB}. We show that, additionally, the turbulent, isotropic form, leads to a correlation between the second derivative of the star's frequency, $\ddot{\nu}$ after a glitch, and the waiting time since the previous glitch (assumed to be a proxy for the velocity lag between superfluid neutrons and normal protons in the pinned superfluid at the glitch epoch). We compare our predictions to measurements of $\ddot{\nu}$ after glitches in the Vela pulsar and in PSR J0537-6910, which are both `regular' glitchers in which we observe a majority of large glitches that occur quasi-periodically \citep{Howitt18}. Although the scatter of the data does not allow for sound statistical conclusions, it is suggestive of a trend for higher values of $\ddot{\nu}$ to occur after longer waiting times, and is compatible with our model given a mutual friction parameter $\mathcal{B}\approx 10^{-6}$. 
This value is consistent with theoretical expectations for phonon mediated mutual friction in the crust \citep{HasSed}. We also find that the turbulent model is a better fit to the relaxations of individual glitches in the Vela pulsar.

Our study thus suggests that the late time relaxations of glitches in the Vela pulsar and in PSR J0537-6910, in which $\ddot{\nu}\approx$ constant \citep{ES17, aek+18}, are due to the response of a pinned, turbulent, superfluid in the crust of the neutron star, and that the glitch itself is triggered in a different region, possibly in the outer core of the star, as suggested by \citet{crab}.
Future high cadence observations of glitch relaxations will allow to either rule out or confirm this model, and obtain constraints on the moment of inertia of the crustal region, and thus also on the mass of the neutron star and its equation of state \citep{hea+15, pah+17}.

{Finally, we suggest that our theoretical description of mutual friction in pinned, turbulent superfluids, could be investigated experimentally by studying systems where superfluid Helium is introduced in porous materials such as aerogels \citep{Aero}. This would allow to have pinning sites in the bulk of the superfluid, and study their effect on turbulence and on the spindown of the container, essentially simulating a neutron star crust in the laboratory.}

\section*{Acknowledgments}
BH and DA acknowledge support from the Polish National Science Centre grant SONATA BIS 2015/18/E/ST9/00577. Partial support comes from PHAROS, COST Action CA16214.

\bibliographystyle{mnras}
\bibliography{crab} 

\appendix
\section{Glitch recoveries data}
\label{appendix}

Post-glitch recoveries of the Vela pulsar are characterised by a strong initial recovery, which can be described by exponentials of multiple timescales, the longest of which is typically around 100 days, followed by a rather linear regime. In order to calculate the second frequency derivative that characterises that late ``linear" part, we perform a linear fit on the (derived) data of the spin-down rate $\dot{\nu}$ that have been presented in \citet{ES17} (e.g. see their figure 3). The dates of glitches used to calculate the inter-glitch intervals ($\Delta T$) were taken from the online glitch database of the Jodrell Bank Centre for Astrophysics (http://www.jb.man.ac.uk/pulsar/glitches.html, \citep{els+11}).

\begin{figure}
\centerline{
 \includegraphics[width=0.5\textwidth]{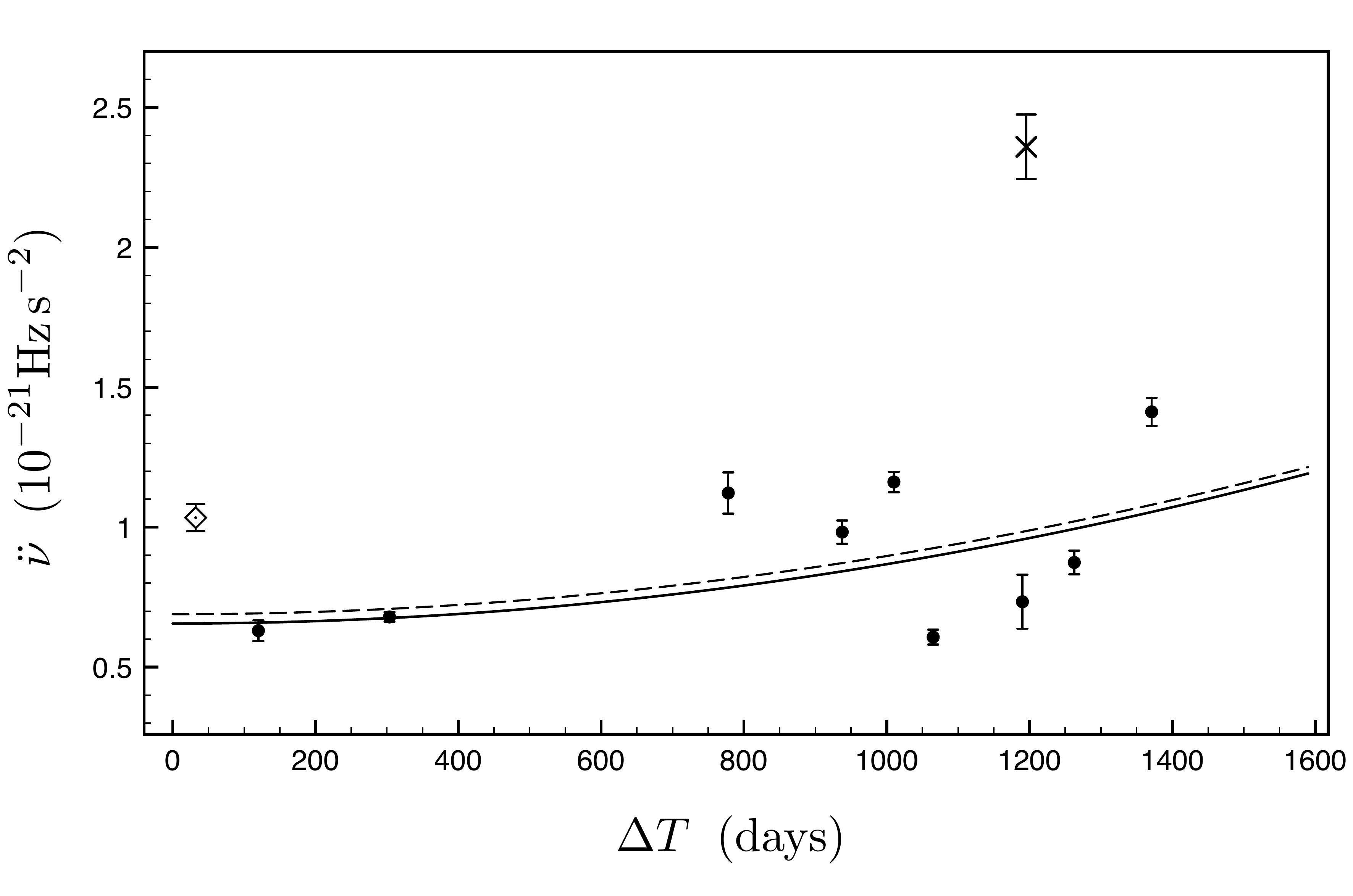}}
 \caption{Measured second frequency derivative $\ddot{\nu}$ for the quasi-linear recovery part after the Vela glitches between 1969 and 2004, plotted versus the inter-glitch waiting time preceding each glitch. Black dots represent $\ddot{\nu}$ measurements from the longer ($>900$ days) inter-glitch intervals, diamond from the 778 days interval and cross for the 303 days interval (see text for details). The dashed line corresponds to the best-fit quadratic function of the form $\ddot{\nu}=a+b t^2$ to all points, whilst the solid line is a fit only to the points from the long intervals. The parameters are $a=0.689$, $b=2.08e-7$  and $a=0.6559$, $b=2.12e-7$ respectively.}
   \label{vela_1}
\end{figure}

\begin{figure}
\centerline{
 \includegraphics[width=0.5\textwidth]{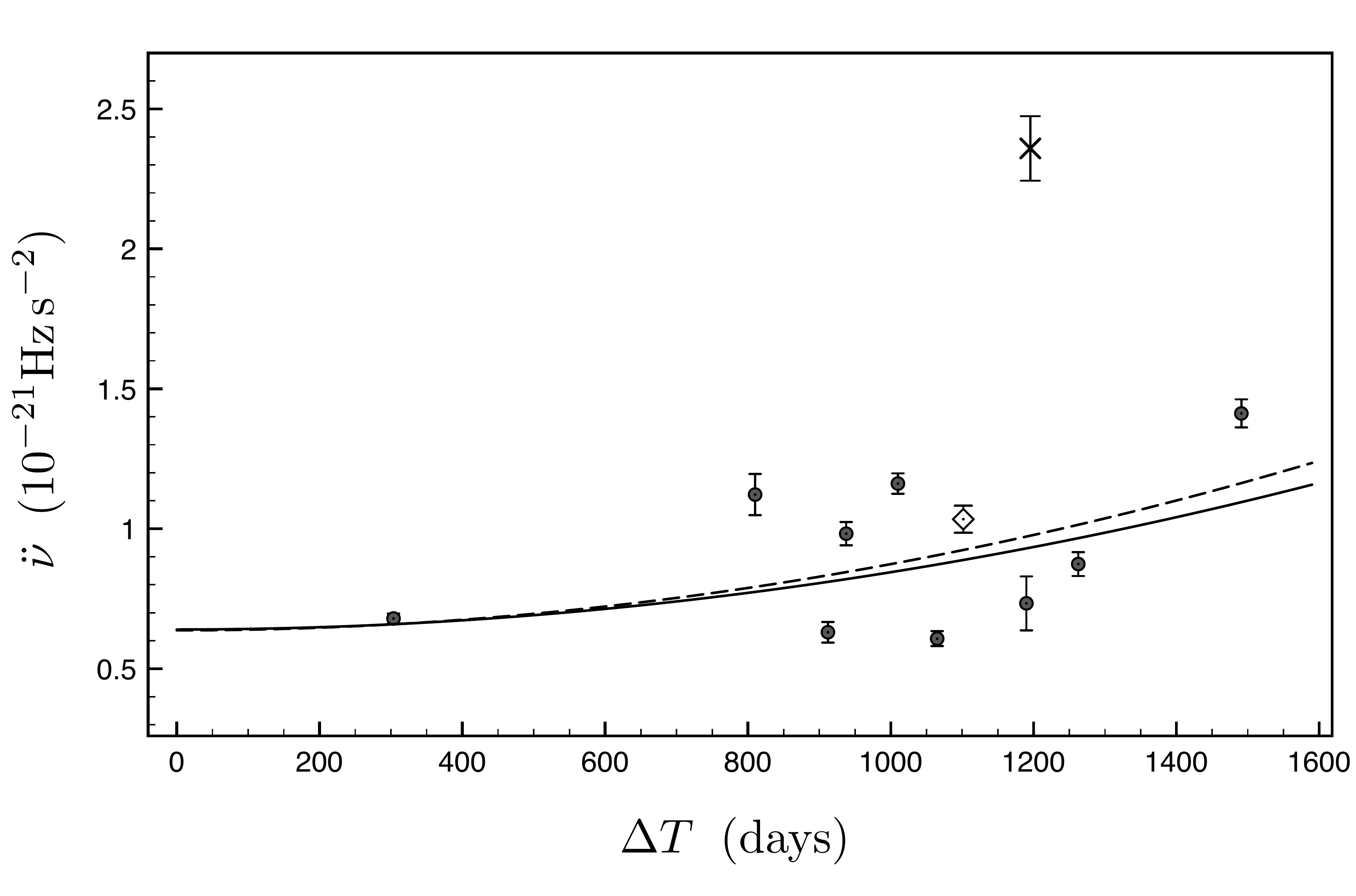}}
 \caption{As in figure \ref{vela_1} but with $\Delta T$ calculated after excluding the two smallest glitches present in the dataset - see text for details.}
   \label{vela_2}
\end{figure}

To avoid a strong influence from the initial recovery phase, we focus on the 10 longest ($> 900$ days) post-glitch time intervals in the $\dot{\nu}$ dataset, and only fit for the last 400 days before the next glitch.  Four inter-glitch intervals do not fulfil the above criterion, however two of them well exceed 100 days, with lengths 303 days and 778 days. For these, we also attempt a linear fit to obtain a constant $\ddot{\nu}$ using only data after 120 days and 250 days post-glitch respectively.  Figure \ref{figvela} displays measurements coming from both the longest intervals and these two short ones. In figure \ref{vela_1} we highlight the two (likely less accurate) measurements that correspond to the shorter intervals and present the best-fit quadratic curves that include or exclude those two points. As can be seen, there is little effect on the best-fit parameters, well within their uncertainties.  
For the years covered by the $\dot{\nu}$ dataset used here, most glitches are of large -typical for the Vela pulsar- size around $\Delta\nu\sim20\,\mathrm{\mu Hz}$ . There are, however, two much smaller events that occurred close to a large glitch (at MJD 413212 and 49591). It is unclear whether those mini glitches will reset the lag in the region that drives the linear recovery, therefore they should perhaps not be considered when calculating the intervals $\Delta T$. We recalculated $\Delta T$ excluding these two events and present the alternative results (and fitted curves) in Figure \ref{vela_2}. Again, the best-fit parameters only vary within uncertainties and there is little effect on the inferred $\mathcal{B}$ parameter. 

\begin{figure}
\centerline{
 \includegraphics[width=0.5\textwidth]{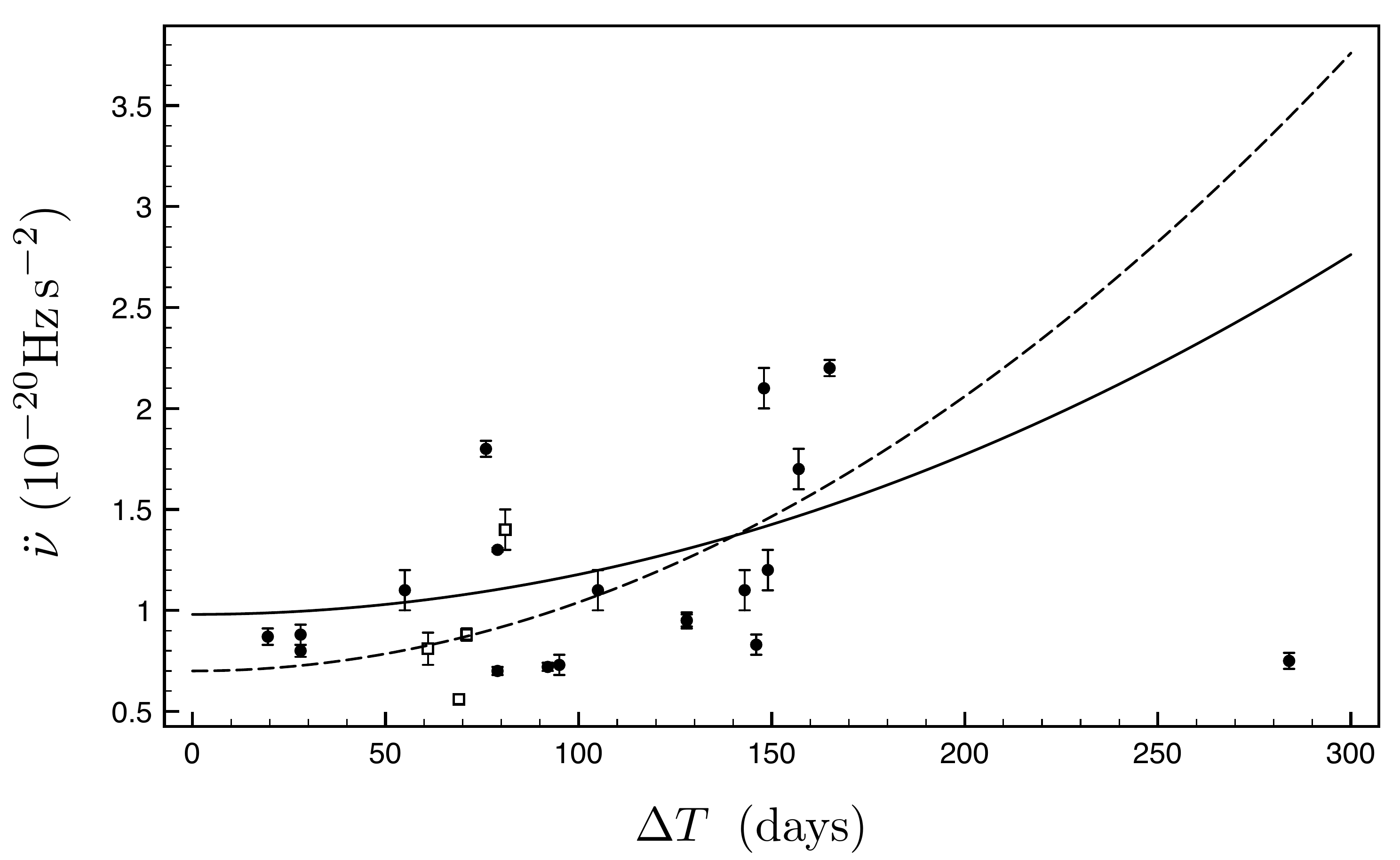}}
 \caption{Measured second frequency derivative $\ddot{\nu}$ between glitches of the pulsar PSR~J0537-6910, plotted versus the inter-glitch waiting time preceding each glitch. Black dots correspond to {\it RXTE} observations from \citet{aek+18}, whilst open squares are the 4 additional datapoints after the new {\it NICER} observations \citep{NicerWynn}. We applied the same selection criteria as in figure \ref{fig0537}. The solid curve is the best fit to the {\it RXTE} data alone, whilst the dashed line is a quadratic ($\ddot{\nu}=a+b t^2$) fit to all the data, with parameters $a=0.7$  and $b=3.4\times 10^{-05}$ days$^{-2}$ (see text for details).}
   \label{nicer}
\end{figure}

In the case of PSR~J0537-6910, we have used the measurements of $\ddot{\nu}$ from \citet{aek+18}, who use all archival {\it RXTE} data. {\it RXTE} was decommissioned on 2012 and the pulsar was not being monitored since. Recently however, in August 2017, {\it NICER} began observations of this source and up to April 2020 has detected an additional 8 glitches. Applying the same selection criteria as in Figure \ref{fig0537} there remain 4 new measurements of $\ddot{\nu}$ and $\Delta T$, as presented in \citet{NicerWynn}, which we incorporate together with the previous data in Figure \ref{nicer}. They follow a similar trend, although their inclusion alters the best-fit parameters of the function $\ddot{\nu}=a+b t^2$ . 
Our main conclusions do not change however, with the newly calculated mutual friction coefficient changing slightly to $\mathcal{B}=(2.1\pm0.4) \times10^{-6}$.
 
\end{document}